\documentclass[12pt,english]{article}
\usepackage[english]{babel}
\usepackage{url}
\usepackage{amssymb,amsmath,amsthm,mathrsfs,latexsym,amsxtra,graphicx,appendix,hyperref,fontenc,cite}

\makeatletter
\renewcommand\section{\@startsection {section}{1}{\z@}%
                                   {-5.5ex \@plus -1ex \@minus -.2ex}
                                   {2.3ex \@plus.2ex}%
                                   {\normalfont\large\bfseries}}
\renewcommand\subsection{\@startsection{subsection}{2}{\z@}%
                                     {-3.25ex\@plus -1ex \@minus -.2ex}%
                                     {1.5ex \@plus .2ex}%
                                     {\normalfont\bfseries}}

\numberwithin{equation}{section}

\makeatother

\newcommand{\bea}{\begin{eqnarray}}
\newcommand{\eea}{\end{eqnarray}}
\newcommand{\be}{\begin{equation}}
\newcommand{\ee}{\end{equation}}

\newcommand{\Z}{{\mathbb Z}}

\newcommand{\C}{{\mathbb C}}

\newcommand{\bH}{{\mathbb H}}

\def\e{\varepsilon}

\newcommand{\cF}{{\cal F }}

\newcommand{\alphab}{{\bar \alpha}}

\newcommand{\ie}{{\it i.e.~}}

\newcommand{\ve}{\vec\varepsilon}

\newcommand{\zb}{{\bar z}}
\newcommand{\xb}{{\bar x}}
\newcommand{\hb}{{\bar h}}
\newcommand{\qb}{{\bar q}}
\newcommand{\Hb}{{\bar H}}

\newcommand{\Greg}{G^{(r)}}

\begin{document}

\begin{titlepage}
	\begin{center}
		
		\hfill \\
		\hfill \\
		\vskip 0.75in
		
		{\Large 
			\bf Conformal Perturbation Theory for Twisted Fields
		}\\

		\vskip 0.4in
		
		{\large Christoph A.~Keller${}^{a}$ and Ida G.~Zadeh${}^{b}$
		}\\
		\vskip 4mm

		${}^{a}$
		{\it Department of Mathematics, University of Arizona, Tucson, AZ 85721-0089, USA} \vskip 1mm
		${}^{b}$
		{\it International Centre for Theoretical Physics, Strada Costiera 11, 34151 Trieste, Italy} \vskip 1mm

		\texttt{cakeller@math.arizona.ch,~zadeh@ictp.it}

	\end{center}
	
	\vskip 0.35in
	
	\begin{center} {\bf ABSTRACT } \end{center}
	We investigate second order conformal perturbation theory for $\Z_2$ orbifolds of conformal field theories in two dimensions. To evaluate the necessary twisted sector correlation functions and their integrals, we map them from the sphere to its torus double cover. We discuss how this relates crossing symmetry to the modular group, and introduce a regularization scheme on the cover that allows to evaluate the integrals numerically. These methods do not require supersymmetry. As an application, we show that in the torus orbifold of 8 and 16 free bosons, $\Z_2$ twist fields  are marginal at first order, but 
	stop being marginal 
	at second order.
	\vfill
	

\end{titlepage}

\tableofcontents

\section{Introduction}

Conformal perturbation theory has a long history \cite{KADANOFF197939, Kadanoff:1978pv,Zamolodchikov:1987ti}. Conceptually, it is straightforward enough: if a conformal field theory (CFT) has an exactly marginal field $\Phi$, then we can obtain a family of theories by perturbing by $\Phi$. If $\Phi$ is exactly marginal, that is if its conformal weight remains unchanged under perturbation, these theories will all be conformal.  Technically, however, perturbation theory is hard. This fact is not apparent at first order, since here the functional form of the 3-point function is completely fixed and determined by a single constant. The integral is thus straightforward to evaluate. An immediate consequence of this is for instance the fact that the shift in the conformal weight of a field $\varphi$ is proportional to its 3-point function coefficient $\langle\varphi\,\varphi\,\Phi\rangle$ \cite{Dijkgraaf:1987jt,Cardy:1989da}. This gives rise to the well-known criterion that a marginal field remains marginal only if its 3-point function with any other marginal field vanishes.

It is only at second order that the real technical difficulties of perturbation theory first become apparent. They arise in two forms: First, the 4-point function that needs to be integrated no longer has a universal form, but is theory dependent. The closest one can come to a universal form is to expand the correlation function in conformal blocks. For non-rational CFTs this expansion has an infinite number of terms, and convergence issues arise. Second, integrals over multiple variables now need to be evaluated. Their regularization thus becomes more subtle. Dealing with these difficulties is the main goal of this paper. We approach them here using tools from theory of modular forms.

Because of the aforementioned difficulties it could be tempting to just stop at first order. In practice, however, there are many questions for which first order is not enough. This often happens when the first order contribution vanishes. A typical example of this involves lifting of fields that saturate some unitarity bound: in that case, their weight is minimal, and can therefore not depend linearly on the perturbation parameter $\lambda$. This is for instance what happened in \cite{Keller:2019suk} and \cite{Gaberdiel:2015uca}, where BPS states and holomorphic states were lifted, respectively, at second order. Second order perturbation theory has also been used to investigate current-current deformations \cite{Ludwig:2002fu,Behr:2013vta} and bulk-boundary interactions \cite{Gaberdiel:2008fn}.

In the context of microstates of black holes in string theory and the D1-D5 brane system \cite{Strominger:1996sh}, there exists a point on the moduli space of the brane system where the theory is described by a $\sigma$-model whose target space is a symmetric product orbifold of K3's or $\mathbb T^4$'s \cite{Vafa:1995bm}. The orbifold point describes the low-energy dynamics of the brane configuration and sits far away from the point associated with the supergravity description. Conformal perturbation theory at second order has been used at the symmetric orbifold point to study properties of states which are protected across the moduli space all the way to the supergravity regime, versus stringy states which are lifted away from the orbifold point \cite{Gava:2002xb,Pakman:2009mi,Burrington:2012yq,Hampton:2018ygz,Guo:2019pzk}.

Non-renormalization theorems for protected quantities of the moduli space of the D1-D5 system such as 3-point functions of BPS states have been developed using conformal perturbation theory and superconformal Ward identities \cite{deBoer:2008ss,Baggio:2012rr}. Moreover, higgsing of the higher spin symmetry generators of the symmetric orbifold theory under the deformation associated with turning on the string tension was analyzed in \cite{Gaberdiel:2015uca}. For applications of conformal perturbation theory in AdS/CFT correspondence in general spacetime dimensions see \cite{Berenstein:2014cia}.

Another application of second order perturbation theory is to compute curvature on the moduli space \cite{Kutasov:1988xb,Friedan:2012hi,Gomis:2016sab}. The perturbing field has to be marginal at least up to second order and the resulting curvature comes from the constant term in the double integral. Deformation of 2d CFTs by chiral irrelevant operators has been studied at second order for CFTs with $\mathcal W$-algebra symmetries \cite{Datta:2014ska,Datta:2014uxa,Datta:2014zpa}. In particular, second order deformation on torus has been studied in \cite{Datta:2014zpa} where a prescription for computing integrals of correlation functions of chiral irrelevant fields is developed.

In this paper, we are interested in a different question: given a marginal field $\Phi$, is it really a modulus, \ie does it remain marginal under perturbations? There is a general belief that this only happens if the CFT is either supersymmetric or free; otherwise, the dimension of $\Phi$ will not be protected at higher orders in perturbation theory, and it will stop being marginal. This question was considered for certain CFTs in \cite{Cardy:1987vr}. In this article, we test this belief for twisted fields in $\Z_2$ orbifolds of free bosons on the torus.

The $\Z_2$ orbifold theory is in a sense the closest to a free theory without actually being free: In the untwisted sector, all correlators are still free boson correlators. In the twisted sector, however, correlation functions are no longer free. We will therefore concentrate on the twisted sector. There the ground state field $\sigma$ has weight
\be
h_\sigma = \left(\frac c{16},\frac c{16}\right)\ .
\ee
For 16 free bosons, $\sigma$ is thus marginal. Similarly, for 8 free bosons, the field $\partial X_{-\frac12} \bar\partial X_{-\frac12} \sigma$ is marginal. These are the marginal operators $\Phi$ that we study. Our goal is to investigate if they remain marginal, or if they become lifted. At first order, there is an immediate answer to this question: due to the twist selection rule, we have \be
\langle\Phi\,\Phi\,\Phi\rangle = 0
\ee
so that $\Phi$ remains indeed marginal. The real question that we need to address thus is if they get lifted at second order. The main result of this work is to show that, perhaps not surprisingly, they do indeed get lifted.

To compute this lifting, it is necessary to integrate a 4-point function. At second order in perturbation theory, we need to evaluate the double integral of a 4-point function. This double integral may be simplified by using a global conformal transformation that maps the insertion points to $0,1,\infty$, and $x$, the cross ratio. Up to regularization issues which we shall discuss later, the double integral then factorizes into two parts: a universal part independent of the correlation function, which gives the logarithmic dependence that leads to the shift in weight, and a part containing the integral over $x$, which gives a constant. This constant multiplies the logarithm and thereby determines the shift in the conformal weight: concretely, the shift at second order in conformal weight of a field $\varphi$ is given by
\be\label{introint}
\int d^2 x \,\langle \varphi(\infty)\Phi(1)\Phi(x)\varphi(0)\rangle\ .
\ee
Note that this integral is divergent and needs to be regularized; we will discuss this in detail in a later section. Assuming no contribution at first order, the shift in conformal weight of $\varphi$ is given by
\be
h(\lambda)= h - \frac{\pi}2 \lambda^2 \int d^2 x \Greg(x) +\ldots\ ,
\ee
where $\Greg(x)$ is a regularized version of the correlation function in (\ref{introint}).

To perform the integral (\ref{introint}), we of course first need to compute the integrand.
In the case at hand, we take $\varphi$ to be the marginal field $\Phi$.  The correlator in (\ref{introint}) is thus a 4-point function of four twisted fields, leading to two $\Z_2$ branch cuts. For our computations, we map the correlator to the double cover of the sphere, which is a torus. This trick works for general 4-point functions \cite{Zamolodchikov1987}, but it is particularly useful here for four twist fields: the most natural way to express the correlator is as a correlator of free bosons on this torus. Moreover, the cover map turns crossing symmetry into symmetry under the modular group $\Gamma_1=SL(2,\Z)$. More precisely, the resulting correlation function on the torus is invariant under a congruence subgroup $\Gamma(2)$ of $\Gamma_1$, and the crossing group is given by $S_3 = \Gamma_1/\Gamma(2)$. The correlation function is then a type of non-holomorphic theta-function in the torus modulus $\tau$, dressed with additional factors.

To evaluate the integral (\ref{introint}),
we note that geometrically, it corresponds to the integral over the moduli space $\hat \C -\{0,1,\infty\}$ of a sphere with four (marked) punctures. On the double cover, the integral then turns into an integral over the moduli space $\bH/\Gamma(2)$. To perform this integral, we first expand the correlation function in $q$ and $\qb$, and then use the Stokes' theorem to reduce the integral to a non-holomorphic contour integral which we evaluate numerically term by term. This method converges very quickly, and from our numerical results we can conclude that $\Phi$ gets lifted at second order.

Perturbation theory of torus orbifolds was studied in \cite{Eberle:2001jq} where the lifting of untwisted momentum states was investigated. That computation is technically much easier, as it only involves two twist fields. The correlation function can therefore be evaluated as an untwisted correlation function on the double cover, which is in this case a sphere. Here we study the lifting of twisted states. This means that the correlation function has four twist fields, so that the cover is a torus. In particular, the correlation function will automatically contain the entire spectrum coming from this torus partition function.

This article is organized in the following way: In section~\ref{s:perThy} we review conformal perturbation theory at first and second order. We discuss how to use conformal transformations to simplify the resulting integrals, and introduce the regularization scheme that we will use. In section~\ref{s:cover} we review the map to the double cover and give explicit expressions for the action of the crossing group and its modular presentation. Section~\ref{s:Integral} explains how to perform the integral on the cover, and how to regularize the integral. Finally, in sections~\ref{s:T16} and \ref{s:T8} we apply our methods to two examples: 16 orbifolded free bosons on a torus, where the twist ground state $\sigma$ is marginal, and 8 orbifolded free bosons, where the field $\partial X_{-\frac12} \bar \partial X_{-\frac12} \sigma$ is marginal. As expected, in both cases we find that the marginal fields are lifted at order two in perturbation theory.

\section{Perturbation theory and regularization}\label{s:perThy}
\subsection{Perturbation theory and integrals on the complex plane}
Perturbation theory is the expansion of expressions such as
\be\label{2ptpert}
\langle\varphi(z_1)\varphi(z_2)\rangle_\lambda = \langle\varphi(z_1)\varphi(z_2) e^{\lambda\int d^2w \Phi(w)}\rangle
\ee
in powers of the coupling $\lambda$.
Here $\Phi$ is a marginal field. Expanding the exponential order by order in $\lambda$, we obtain integrals that need to be suitably regularized.
The correlator (\ref{2ptpert}) leads to a shift of the weight $\Delta$ of $\varphi$. This shift occurs if perturbation theory produces logarithmic terms $\log|z_1-z_2|$. Because of dimensional analysis, such terms will of course always be accompanied by logarithmic terms in the regularization parameter $\epsilon$.

Evaluating (\ref{2ptpert}) involves integrating correlation functions over the complex plane, with various discs cut out for regularization purposes.
There are various tricks to evaluate such integrals. The most basic one is to apply Stokes' theorem in complex coordinates,
\be
\int_{\partial U} F dz + G d\zb = \int_U \left(\partial_z G- \partial_\zb F\right) dz d\zb
\ee
where the complex integration measure is given by
\be
dx\wedge dy = \frac{i}{2} dz \wedge d\bar z\ .
\ee
If the integrand happens to be a total derivative, then this turns the integral into a contour integral around the $\epsilon$-discs. This happens for instance if there are Ward identities for supersymmetric correlation functions such as in \cite{Keller:2019suk}. For fields that are not sufficiently protected by supersymmetry, or for theories that are not supersymmetric at all, we cannot expect this to happen.

The next best case is if the integrand factorizes into a holomorphic and an anti-holomorphic part. One can then take an anti-derivative of one of them and apply Stokes again. However, in such cases it often happens that the anti-derivative introduces branch cuts. One can then either carefully evaluate the contour integral including those cuts, or alternatively use what are called `Riemann bilinear relations'\footnote{We thank Hirosi Ooguri for pointing these out to us.}; their application is for instance described in \cite{Kawai:1985xq,MR972993}.

In general however the correlation functions will not factorize. The typical case for second order perturbation theory is to integrate a 4-point function over its cross section,
\be
G(x,\xb)= \sum_{\phi}C_{12\phi}C_{34\phi}\cF_{\phi}(x)\cF_\phi(\xb)\ .
\ee
This is exactly the case that we encounter. We then expand $G$ and integrate it term by term up to a certain cutoff, giving us an approximate numerical result.

\subsection{First order perturbation theory}
Let us first review perturbation theory at first order. In that case we have
\be
\delta \langle\varphi(z_1)\varphi(z_2)\rangle
= \lambda \int d^2 w \langle \varphi(z_1)\varphi(z_2)\Phi(w)\rangle_{reg}\ .
\ee
We want to use a hard sphere cutoff regularization scheme: that is, we cut out discs of radius $\epsilon$ around the insertion points $z_{1,2}$. The 3-point function has the universal form
\be
\langle \varphi(z_1)\varphi(z_2)\Phi(w)\rangle
= \frac{C_{\varphi\varphi\Phi}}{|w-z_1|^2|w-z_2|^2 z_{12}^{2h-1}\zb_{12}^{2\hb-1}}\ .
\ee

To perform the integral, we want to use various conformal transformations. We first shift $w$ by $z_2$, and then perform the coordinate transformation determined in \cite{Eberle:2001jq}:
\be\label{eberle_i}
x(w)=\frac{z_{12}w}{z_{12}-w}\ ,
\ee
giving
\be\label{phi_phi_O_ii}
\frac{C_{\varphi\varphi\Phi}}{z_{12}^{2h}\;\zb_{12}^{2\bar h}}\int\frac{d^2x}{x\bar x}\ .
\ee
We have to be careful about the shape of the cut-out disks under the $SL(2,\mathbb C)$ transformation (\ref{eberle_i}): the regularization scheme changes in that the shape of the hard spheres are modified at subleading order in the cut-off radius. We therefore need to take into account the subleading terms for the squashed spheres.
More precisely, for the small circles with radius $\epsilon$ cut out around $w=0$ and $w=z_{12}$, eq. (\ref{eberle_i}) gives
\bea\label{disks_y}
&&x\big(w=\epsilon e^{i\theta}\big)=e^{i\theta}\epsilon+\frac{e^{2i\theta}}{z_{12}}\,\epsilon^2+\frac{e^{3i\theta}}{z_{12}^2}\,\epsilon^3+\cdots\ ,\\
&&x\big(w=z_{12}+\epsilon e^{i\theta}\big)=-\frac{z_{12}^2}{\epsilon e^{i\theta}}-z_{12}\ .
\eea

We can then evaluate (\ref{phi_phi_O_ii}) order by order in $\epsilon$, for instance by going to polar coordinates and keeping track of the boundaries of the squashed discs around 0 and $\infty$. This gives
\be\label{int_phi_phi_O_i}
\frac{2\pi C_{\varphi\varphi\Phi}}{z_{12}^{2h}\;\zb_{12}^{2\bar h}}\;\ln\left(\frac{|z_{12}|^2}{\epsilon^2}\right) + o(1)\ .
\ee
To cancel the divergent part, we need to insert a counterterm  
\be
2\pi C_{\varphi\varphi\Phi} \lambda \log \epsilon^2\ .
\ee
The logarithmic term then leads to the shift of the conformal weight,
\begin{multline}
\;\qquad\langle\varphi(z_1)\varphi(z_2)\rangle_\lambda
= \frac1{z_{12}^{2h}\;\zb_{12}^{2\bar h}}(1+2\pi\lambda  C_{\varphi\varphi\Phi}\log(|z_{12}|^2)+O(\lambda^2))\\
= \frac1{z_{12}^{2h-2\lambda\pi C_{\varphi\varphi\Phi}}\;\zb_{12}^{2\hb-\lambda2\pi C_{\varphi\varphi\Phi}}} + O(\lambda^2)
= \frac1{z_{12}^{2h(\lambda)}\;\zb_{12}^{2\hb(\lambda)}} + O(\lambda^2)\ .
\end{multline}

We then obtain
\be\label{anomdim_1_i}
h(\lambda)= h - \pi C_{\varphi\varphi\Phi}\lambda + O(\lambda^2)
\ee
and similar for $\hb(\lambda)$, which reproduces indeed the expected result for the shift of the conformal dimension \cite{Cardy:1989da}. Moreover, note that (\ref{int_phi_phi_O_i}) shows that our hard sphere regularization has not produced a constant term. A constant term would have led to the introduction of Christoffel symbols for the curvature of the moduli space \cite{Kutasov:1988xb}. This confirms the remark in \cite{Friedan:2012hi} about hard sphere regularization not introducing any Christoffel symbols.

\subsection{Second order perturbation theory}\label{s:2ndorder}

Let us now move on to second order perturbation theory. To this end, we briefly repeat the discussion in \cite{Keller:2019suk}. The second order term in perturbation theory reads
\be\label{2ndorderint}
\frac{\lambda^2}{2} \int d^2 w_1 d^2 w_2 \langle \varphi(z_1)\Phi(w_1)\Phi(w_2)\varphi(z_2)\rangle\ .
\ee
This integral is again divergent and needs to be regularized both in the UV and IR. We will ignore this issue for a moment, and return to it later. Just as before, we want to use global conformal symmetry to simplify the integral: the M\"obius transformation
\be\label{mobius}
z \mapsto f(z):= \frac{(z-z_2)(w_1-z_1)}{(z-z_1)(w_1-z_2)}
\ee
allows us to rewrite the expressions in terms of the cross-ratio $x:= f(w_2)$, which we use to replace $w_2$. It also changes the integration measure to
\be
d^2 w_1 d^2 w_2 = d^2 w_1 d^2 x 
\left| \frac{\partial(w_1,w_2)}{\partial(w_1,x)}\right|^2 
=d^2 w_1 d^2 x \left|\frac{(z_1-w_2)^2 (w_1-z_2)}{(z_1-w_1) (z_1-z_2)} \right|^2
\ee
so that the integral (\ref{2ndorderint}) turns into
\be\label{2ndorderint2}
\frac{\lambda^2}{2}\int d^2 w_1\,z_{12}^{-2h_\varphi}\zb_{12}^{-2\bar h_\varphi}\left|\frac{z_1-z_2}{(z_1-w_1) (w_1-z_2)} \right|^2
\int d^2 x \,\langle \varphi(\infty)\Phi(1)\Phi(x)\varphi(0)\rangle\ .
\ee

The $x$ integral now seems to be independent of $w_1$ and so we could just naively evaluate the $w_1$ integral. This integral is divergent in the UV but converges at $w_1=\infty$. We therefore need to regularize it through cutting out $\epsilon$-discs around $z_1$ and $z_2$. We find
\be\label{2ndorder}
\pi\lambda^2 \log\left(\frac{|z_{12}|^2}{\epsilon^2}\right) z_{12}^{-2h_\varphi}\zb_{12}^{-2\bar h_\varphi}
\int d^2 x \langle \varphi(\infty)\Phi(1)\Phi(x)\varphi(0)\rangle\ .
\ee
The numerical coefficient of $\log |z_{12}|$ is the anomalous dimension of $\varphi$ \cite{Dijkgraaf:1987jt,Cardy:1986ie,Eberle:2001jq}, which is thus given by the $x$ integral of the 4-point function. The problematic issue here is that the $x$ integral is divergent and so we need to regularize it. Naively, it seems that changing the regularization scheme may change the constant part of the integral. This then implies that the anomalous dimension is scheme-dependent. If the first order contribution to the shift vanishes, then this would contradict the general principle that the leading order shift should be scheme-independent. This apparent contradiction shows that we have been too naive.

Let us therefore give a more careful analysis. Our claim is that (\ref{2ndorder}) gives the correct result if we replace the integrand by a suitably regularized 4-point function, which we give in (\ref{H0}). To obtain this result, we have to be careful in computing the $x$ integral. Contrary to our statement above, our regularization scheme in fact \emph{does} introduce a $w_1$ dependence for the $x$ integral. To see this, we note that we need to regularize the integrals in (\ref{2ndorderint}) already. As described above, we regularize the $w_2$ integral by cutting out $\epsilon$-discs around $z_1,w_1,$ and $z_2$, and similar for the $w_1$ integral. In this we need to be careful when both $w_1$ and $w_2$ collide at the same time with another insertion, so that the discs overlap. Moreover we need to introduce an IR regulator to deal with the divergence at infinity. In the following we neglect these two effects, as we believe that they will not affect the final result of our analysis for the following reasons: We apply a conformal transformation that effectively eliminates one of the integrals, so that triple collisions do not happen. This transformation also moves one of the insertions to infinity, so that the UV regulator takes care of the IR regularization as well. It should be possible confirm this belief by a more careful analysis involving the $\beta$ function along the lines of \cite{Gaberdiel:2008fn,Behr:2013vta}, but we have not attempted to do so.

Returning to the apparent contradiction above, the point is that just as with the transformation (\ref{eberle_i}) at first order, (\ref{mobius}) changes the shape of those discs. In particular, the shape and size now also depend on all other insertion points. To make this clearer, let us concentrate on the disc around $z_2$ first. The coordinate transformation (\ref{mobius}) maps it to a disc in the $x$ integral around $x=0$. The cross-ratio is given by
\be
x= \frac{(w_2-z_2)(w_1-z_1)}{(w_2-z_1)(w_1-z_2)}
\ee
so that the boundary of the $\epsilon$-disc of $w_2(\theta) = z_2 +\epsilon e^{i\theta}$ turns into
\be\label{ediscx}
x(\theta) = \frac{w_1-z_1}{w_1-z_2} \frac{\epsilon e^{i\theta}}{z_2-z_1} R(\theta, z_1-z_2)\ ,
\ee
where $R(\theta,z_1-z_2)= 1 + O(\epsilon)$ is independent of $w_1$. As pointed out above, the boundary of the $x$ integral, and therefore also the integral itself, now indeed depend on $w_1$. That is,
\be
M(\epsilon):=\int_{\C-D_\epsilon} d^2 x G(x)\ ,
\ee
where
\be\label{Gx}
G(x):=\langle \varphi(\infty)\Phi(1)\Phi(x)\varphi(0)\rangle\ .
\ee
$D_\epsilon$ is given by eq. (\ref{ediscx}) and is not a round disk in the $x$-space. We note that , actually depends on $w_1$, since $D_\epsilon$ does. We, therefore, cannot simply evaluate the $w_1$ and the $x$ integral independently.

To deal with this issue, let us replace our regularization scheme with a different one. We introduce a regularized four point function
\be
\Greg(x)= \langle \varphi(\infty)\Phi(1)\Phi(x)\varphi(0)\rangle - G_{reg}(x)
\ee
where $G_{reg}(x)$ contains all fields that lead to a divergent integral.
We then have
\begin{multline}\label{Dint}
\qquad\qquad\qquad\qquad\; M(\epsilon)= \int_{\C-D_\epsilon} d^2 x \Greg(x)
+ \int_{\C-D_\epsilon} d^2 x G_{reg}(x)\\
=\int_{\C} d^2 x \Greg(x)
+ \int_{\C-D_\epsilon} d^2 x G_{reg}(x)-\int_{D_\epsilon} d^2 x \Greg(x)\ .
\end{multline}

Our claim is now that only the first integral in the second line of (\ref{Dint}), which is an  $\epsilon$ and $w_1$-independent constant, contributes to the shift in weight.
To establish this, we will only discuss the singularity at $x=0$, as the argument for the other two singularities follows from crossing transformations. The regulator $G_{reg}(x)$ is the sum of regulators at the three singularities. The $x=0$ regulator is given by
\bea\label{G0reg_ii}
G_0(x)&=& \sum_{\Delta_\phi<\Delta_\varphi} \frac{(C_{\Phi\varphi\phi})^2}{x^{1+h_\varphi-h_\phi}\xb^{1+\hb_\varphi-\hb_\phi}}\ .
\eea
We can therefore expand
\be\label{Hexpansion}
\Greg(x) = \sum_{\Delta_\phi\geq \Delta_\varphi}  \frac{g_{h,\hb}}{x^{1+h_\varphi-h_\phi}\xb^{1+\hb_\varphi-\hb_\phi}}\ .
\ee
Defining $a:=\Delta_\phi-\Delta_\varphi$, the integral of a term in (\ref{Hexpansion}) gives
\be\label{Gterm}
\sim \frac{\epsilon^a}{|z_1-z_2|^a} \left|\frac{w_1-z_1}{w_1-z_2}  \right|^a (1+ O(\epsilon))\ ,
\ee
where the higher order correction comes from $R(\theta,z_1-z_2)$ in (\ref{ediscx}), and $a\geq0$ since we removed all fields that lead to a singularity. We note that if $\phi$ and $\varphi$ have different spins, then the $x$-integral vanishes to leading order in $\epsilon$.

We can then perform the integration (\ref{2ndorderint2}) of a term (\ref{Gterm}) with $a>0$. The leading order gives
\be\label{aterm}
\epsilon^{a} 
|z_2-z_1|^{-2\Delta_\varphi+2-a} \int_{D_\epsilon} d^2w_1 
\frac{|z_1-w_1|^{a-2}}{ |w_1-z_2|^{a+2}}
\sim \frac{\epsilon^{2a}}{|z_1-z_2|^{-2\Delta_\varphi +2a}}+ O(\epsilon^{0})\ .
\ee
where the $\epsilon^{2a}$ comes from the singularity at $z_1$, and the $O(\epsilon^0)$ term comes from the singularity at $z_2$. The subleading terms in (\ref{Gterm}) only give higher powers in $\epsilon$. It follows that if $a>0$, $M^a$ does not produce a $\log|z_1-z_2|$ term, and therefore, does not contribute to the shift of the weight. Moreover, in the case we are interested in, we can also exclude the case $a=0$, that is the case $\Delta_\phi=\Delta_\varphi$. Note that the integral leading to eq. (\ref{Gterm}) vanishes to leading order in $\epsilon$ unless $\varphi$ and $\phi$ have the same spin. This means that $a=0$ can only arise if $h_\varphi = h_\phi$ and $\hb_\varphi = \hb_\phi$. In our applications we will take $\varphi$ to be marginal field that remains marginal at first order. This means that there is no marginal field in the OPE of $\varphi$ with $\Phi$, which excludes the case $a=0$. Similarly, we can deal with the singularities at $x=\infty$ and $x=1$. In summary, this establishes that the last integral in (\ref{Dint}) does not contribute to the shift in conformal weight.

The second integral $\int_{\C-D_\epsilon} d^2 x G_{reg}(x)$ in (\ref{Dint}) can in principle make a contribution to the lifting. This happens when its $\epsilon$ expansion has a constant term. In practice $G_{reg}(x)$ is of the simple form (\ref{G0reg_ii}), so that it is straightforward to evaluate this contribution explicitly. In the cases we study we will find that there is in fact no such constant contribution. This establishes that only the first integral in (\ref{Dint}) contributes to the shift in weight.

To summarize, the proper regularization scheme is applied to the $w_1$ and $w_2$ integrals (\ref{2ndorderint}). We define it by cutting out discs of radius $\epsilon$ for $w_1$ and $w_2$. A local regularization scheme requires that the discs are independent of other variables. Transforming to the $x$ integral, we then have the discs in $x$ co-ordinates depending on $z_i$. If we chose to rather define an ad hoc regularization scheme for the $x$ integral through cutting out discs of radius $\epsilon$, we would have found that the scheme is in fact not local. This is because the associated discs in the original co-ordinates will depend on other insertions.

In total, the outcome is the following: The second order contribution to the shift in the conformal dimension is given by
\be
\lambda^2\pi \log(|z_1-z_2|^2) (z_1-z_2)^{-2h_\varphi}(\bar z_1-\bar z_2)^{-2\bar h_\varphi} M^0
\ee
where $M^0$ is the $\epsilon^0$ term in the expansion $M(\epsilon) = \sum_a M^a \epsilon^a$ of (\ref{Dint}). If the regulator does not introduce a constant term, then $M^0$ is given by the integral of the regulated 4-point function:
\be\label{H0}
M^0 = \int_{\C} d^2 x \Greg(x)\ .
\ee
As long as there is no first order contribution (which is the case for the theories we are considering), the shift in the conformal dimension of $\varphi$ is thus
\be\label{hshift2}
h(\lambda)= h - \frac{\pi}2 M^0 \lambda^2+O(\lambda^3)\ .
\ee

We end this section by noting that the integral (\ref{H0}) can be understood as a string theory scattering amplitude. One can write the perturbation theory in terms of the $\beta$ function of the coupling constants $\lambda^i$. In this context, the subtraction of the relevant operators are associated with the subtraction of power-like divergences of the scattering amplitude. It would be interesting to derive a general formula for the $\beta$ function along the lines of the analysis of \cite{Behr:2013vta}.\footnote{We thank the referee for pointing this out to us.}

\section{Crossing symmetry and the cover map}\label{s:cover}

\subsection{Cover map}
To evaluate our 4-point functions, it is useful to map them to the double cover of the sphere, which is a torus. The cross ratio $x$ of the sphere then turns into the modulus $\tau$ of the torus, and the crossing (or anharmonic) group turns into the modular group. In principle, one can use this trick for any 4-point function \cite{Zamolodchikov1987}. It was used for instance in \cite{Maloney:2016kee} to construct a modular sum of conformal blocks to obtain crossing symmetric correlators. In our case, however, this approach will be even more powerful: our correlators have four $\Z_2$ twist fields with two $\Z_2$ branch cuts between pairs of insertions. The double cover is then indeed a torus without any branch cuts, on which we simply compute the correlators of a free theory. 

Let us now give the explicit expressions for the quantities involved.
We start with a Riemann sphere with punctures at $0,1,x,\infty$, and its double cover, the torus with modulus $\tau$. Let $z\in \hat \C$ be the coordinate on the sphere, and $t\in \C/(\Z+\tau\Z)$ be the coordinate on the double cover. 
The cover map is given by the Weierstrass $\wp$ function
\be\label{Weierstrass}
z(t)= \frac{\wp(t)-e_1}{e_2-e_1}\ .
\ee
Here $\wp(t)$ is doubly periodic with periods $1$ and $\tau$, and has a double pole at $t=0$.
The $e_i$ are the values of $\wp$ at the half periods. They can be expressed explicitly
\be\label{ehalfperiod}
e_1 =\wp\Big(\frac12\Big)= \frac{\pi^2}{3}(\theta_3^4+\theta_4^4)\ , \quad
e_2 =\wp\Big(\frac{\tau}{2}\Big)= \frac{\pi^2}{3}(-\theta_2^4-\theta_3^4)\ , \quad
e_3 =\wp\Big(\frac{1}{2}+\frac{\tau}{2}\Big)= \frac{\pi^2}{3}(\theta_2^4-\theta_4^4)\ ,
\ee
where the Jacobi theta functions $\theta_i(z|\tau)$ are given by 
\bea
\theta_1(z|\tau)&=& -i\sum_{n\in\Z}(-1)^nq^{\frac12(n+\frac12)^2} y^{n+\frac12}\ ,\\
\theta_2(z|\tau)&=& \sum_{n\in\Z}q^{\frac12(n+\frac12)^2} y^{n+\frac12}\ ,\\
\theta_3(z|\tau)&=& \sum_{n\in\Z}q^{\frac12n^2} y^{n}\ ,\\
\theta_4(z|\tau)&=& \sum_{n\in\Z} (-1)^n y^n q^{\frac12n^2}\ ,
\eea
with the usual definitions $q=e^{2\pi i\tau}$, $y=e^{2\pi iz}$. In (\ref{ehalfperiod}) and in the following use the convention that any $\theta_i$ written without an argument is the specialized theta function $\theta_i(0|\tau)$.
We will also use the identities
\be
e_1-e_2 = \pi^2 \theta_3^4\ ,\qquad e_1-e_3= \pi^2 \theta_4^4\ , 
\qquad e_3-e_2 = \pi^2 \theta_2^4\ .
\ee
and
\be
e_1+e_2+e_3=0\ ,
\ee
which follows from $\theta_4^4 = \theta_3^4 - \theta_2^4$.
From (\ref{Weierstrass}) it follows that the four punctures map to the torus as 
\be
z\Big(\frac12\Big)= 0\ ,\qquad z\Big(\frac{\tau}2\Big)= 1\ , \qquad z(0) =\infty\ ,
\ee
and 
\be\label{x}
x:= z\Big(\frac{1}{2}+\frac{\tau}{2}\Big) = \frac{\theta_4^4}{\theta_3^4}
= \prod_{n=1}^\infty \left(\frac{1-q^{n-\frac12}}{1+q^{n-\frac12}}\right)^8=
1 - 16 q^{\frac12} + 128 q - 704 q^{\frac32} +\ldots\ .
\ee
In particular this relates the cross ratio $x$ of the sphere to the modulus $\tau$ of the cover torus. It will be necessary to understand the expansion of $z$ around those four points. To this end, we use the fact that the Weierstrass function $\wp$ satisfies the differential equations \cite{Apostol1990}
\be\label{wpprime}
(\wp'(z))^2 = 4(\wp(z)-e_1)(\wp(z)-e_2)(\wp(z)-e_3)
\ee
and
\be
\wp''(z) = 6\wp(z)^2 - (e_1^2+e_2^2+e_3^2)\ .
\ee
From this we can extract
the first two derivatives of $\wp$ at the half-periods,
\begin{align}
\wp\Big(\frac{1}{2}\Big)=e_1\ ,&& \wp'\Big(\frac{1}{2}\Big)=0\ ,&& \wp''\Big(\frac{1}{2}\Big)= 2(e_1-e_2)(e_1-e_3)\ ,\\
\wp\Big(\frac{\tau}{2}\Big)= e_2\ ,&& \wp'\Big(\frac{\tau}{2}\Big)=0\ ,&&\wp''\Big(\frac{\tau}{2}\Big)= 2 (e_2 - e_1) (e_2 - e_3)\ ,\\
\wp\Big(\frac{1}{2}+\frac{\tau}{2}\Big)= e_3\ ,&& \wp'\Big(\frac{1}{2}+\frac{\tau}{2}\Big)=0\ ,&&
\wp''\Big(\frac{1}{2}+\frac{\tau}{2}\Big)= 2 (e_3 - e_2) (e_3 - e_1)\ ,
\end{align}
which give the Taylor expansions
\bea
t=0 :&& z =\frac{1}{e_2-e_1}\,\frac{1}{t^2}+\ldots\ ,\\
t=\frac{1}{2}:&& z=(e_3-e_1)\Big(t-\frac{1}{2}\Big)^2+\ldots\ ,\\
t=\frac{\tau}{2}:&& z=1+(e_2-e_3)\Big(t-\frac{\tau}{2}\Big)^2+\ldots\ ,\\
t= \frac{1}{2}+\frac{\tau}{2}:&& z= x+\frac{(e_3 - e_2) (e_3 - e_1)}{e_2-e_1}\Big(t- \frac{1}{2}-\frac{\tau}{2}\Big)^2+\ldots\ .
\eea
We will make use of these expansions when computing correlators.

\subsection{Crossing symmetry and modular transformations}
\begin{figure}
	\centering
		\includegraphics[width=.8\textwidth]{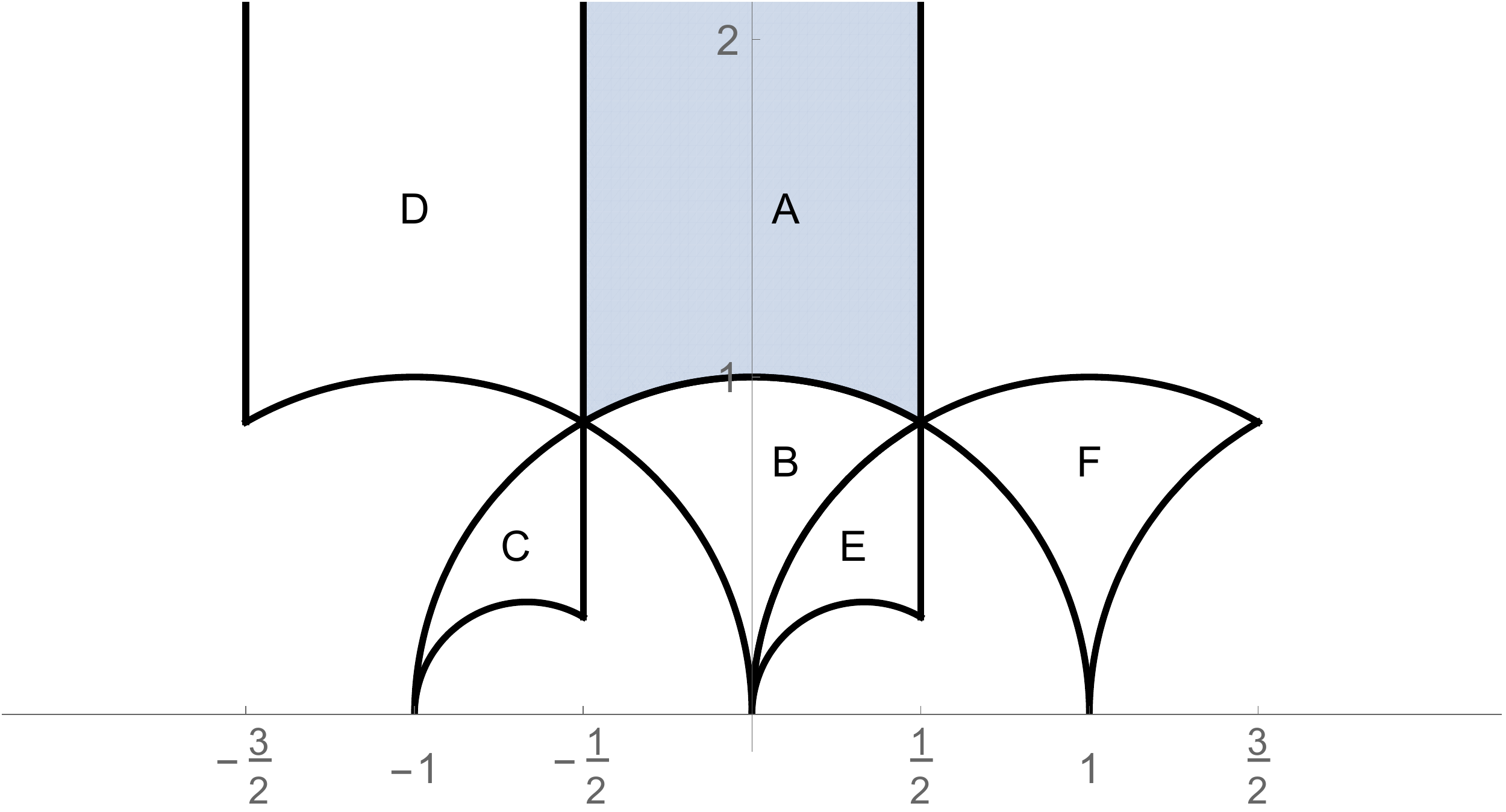}
		\caption{A somewhat non-standard choice for the fundamental region $\cF_2=\bH/\Gamma(2)$. The shaded region denoted by $A$ is the standard `keyhole' fundamental region $\cF_1=\bH/\Gamma_1$ of the full modular group $\Gamma_1$. The other components of $\cF_2$ are the images of $\cF_1$ under the non-trivial elements $B,C,\ldots,F$ of the crossing group $S_3$.}	\label{f:F2} 
\end{figure}

Next let us discuss the action of the crossing group $S_3$, and how it relates to transformations of $\tau$ under the modular group $\Gamma_1$. 
The cross ratio $x(\tau)$ is closely related to the \emph{modular lambda} function $\lambda(\tau)$, 
which is usually defined as
\be
\lambda(\tau)= \frac{\theta_2^4}{\theta_3^4} = 1-x(\tau)\ .
\ee
The modular lambda function, and therefore also $x(\tau)$, is
a Hauptmodul for the congruence subgroup $\Gamma(2)$ given by
\be
\Gamma(2)=\left\{\begin{pmatrix}a&b\\ c&d \end{pmatrix}\in \Gamma_1 : \begin{pmatrix}a&b\\ c&d \end{pmatrix} \equiv \begin{pmatrix}1&0\\ 0&1 \end{pmatrix} \mod 2  \right\}\ .
\ee
Being a Hauptmodul means that $x(\tau)$ is invariant under $\Gamma(2)$, and that it is a bijection from $\C$ to the quotient $\bH/\Gamma(2)$ of the upper half plane by $\Gamma(2)$. A particular choice $\cF_2\subset\bH$ for this quotient, also called a fundamental region, is plotted in figure~\ref{f:F2}.
It has three so-called cusps at $0,1,$ and $i\infty$. These cusps are mapped by $x(\tau)$ to the points
\be\label{cuspmap}
\tau =i\infty \longleftrightarrow x=1\ , \qquad \tau=0 \longleftrightarrow x=0\ ,
\qquad \tau =1 \longleftrightarrow x=\infty\ .
\ee
Note that $x(\tau)$ is invariant under $\Gamma(2)$, but not invariant under the full modular group $\Gamma_1$, which is generated by $T$ and $S$ transformations:
\be\label{SandT}
T: x(\tau+1)= \frac{1}{x(\tau)}\ , \qquad S: x(-{\textstyle\frac{1}{\tau}})= 1-x(\tau)\ .
\ee
For future reference, we invert (\ref{x}) to obtain the expansion of $q$ for $x$ around 1,
\be\label{qexpansion}
q = \frac{(1-x)^2}{256}+\frac{(1-x)^3}{256}+\ldots
\ee
Similarly, it is also useful to define its $S$ transform
\be\label{qtildeexpansion}
\tilde q = q(-{\textstyle\frac1\tau})= \frac{x^2}{256}+\frac{x^3}{256}+\frac{29 x^4}{8192}+\ldots
\ee
which has the advantage that $\tilde q\to0$ as $x\to 0$.

From the above, $x(\tau)$ is invariant under $\Gamma(2)$, and, because it is a Hauptmodul, maps $\C$ one to one to the fundamental region $\cF_2=\bH/\Gamma(2)$ plotted in figure~\ref{f:F2}. Note that of course any function of $x(\tau)$ is automatically invariant under $\Gamma(2)$, but not necessarily invariant under $\Gamma_1$. In fact, this is exactly where the crossing (or anharmonic) group appears: $\Gamma(2)$ is an index 6 subgroup of $\Gamma_1$, and in fact we have
\be
S_3\cong\Gamma_1/\Gamma(2)
\ee
which is the crossing group. We will label its six elements by $\gamma=A,B,C,D,E,F$. Note that it is generated by the two transformations (\ref{SandT}), $S$ and $T$.

The crossing group acts by permutation on the four fields. To illustrate this, consider the correlation function of four marginal fields $\Phi_i$,
\be
G_{\bf 1}(x):= \langle \Phi_{1}(\infty)\Phi_{2}(1)
\Phi_{3}(x)
\Phi_{4}(0) \rangle
\ee
where ${\bf 1}$ denotes the identity element of the crossing group $S_3$. The crossing groups acts on the fields by permuting the indices $124$. This means that if we take some element $\gamma \in \Gamma_1/\Gamma(2)$, then
\be\label{Gcross1}
G_{\bf 1}(x) \sim G_{\sigma_\gamma}(\gamma x) \ ,
\ee
where $\sigma_\gamma$ permutes the the fields 1,2, and 4. The $\sim$ in (\ref{Gcross1}) comes from the fact that under some of the crossing transformations, we pick up an additional $x$-dependent prefactor. For instance
\be\label{G4ptcrossing}
G_{\bf 1}(1-x)=G_{(24)}(x)\ , \qquad G_{\bf 1}({\textstyle\frac1x}) =|x|^4G_{(14)}(x)\ .
\ee
Since these two transformations generate the crossing group, we can in principle obtain all prefactors from (\ref{G4ptcrossing}). Instead, we prefer to define a correlation function in the torus variables
\be\label{Fdef}
F_\sigma(\tau):= y^2 \left|\frac{\partial x}{\partial\tau}\right|^2G_{\sigma}(x)
\ee
Here, $\left|\frac{\partial x}{\partial\tau}\right|^2$ is the Jacobian of the cover map which has  $q$-expansion
\be\label{JacExp}
\left|\frac{\partial x}{\partial\tau}\right|^2 = 4\pi^2( 64 q^{\frac12}\qb^{\frac12}+\ldots)
\ee
and $y$ is the imaginary part of the modulus $\tau = x +iy$.

We will see that $F_\sigma(\tau)$ is in fact the integrand that appears in the integral on the cover. For the moment we simply note that $F_\sigma(\tau)$ has nicer transformation properties than $G_\sigma(x)$ because the Jacobian absorbs the prefactors: First note that $F_\sigma(\tau)$ is invariant under $\Gamma(2)$, since $x$ is invariant under $\Gamma(2)$, and the $\tau$-derivative and $y$ together are invariant under $\Gamma_1$. Second, note that
\be
F_{\bf1}(\tau+1)= y^2 \left| \frac{\partial x^{-1}}{\partial \tau}\right|^2G_{\bf 1}({\textstyle\frac1x})= y^2 |x'|^2 |x|^{-4}G_{(14)}(x)|x|^4= F_{(14)}(\tau)
\ee
and
\be
F_{\bf 1}(-{\textstyle\frac1\tau})= \frac{y^2}{|\tau|^4}|\tau|^4\left| \frac{\partial (1-x)}{\partial \tau}\right|^2G_{\bf 1}(1-x)=  y^2 |x'|^2 G_{(24)}(x)
= F_{(24)}(\tau)\ .
\ee
Since these two generate $S_3$, and $F(\tau)$ is invariant under $\Gamma(2)$, it follows that
\be\label{Ftrafo}
F_{\bf 1} (\gamma \tau) = F_{\sigma_\gamma }(\tau)\ .
\ee
Comparing this to (\ref{Gcross1}), we see that in (\ref{Ftrafo}) we could absorb all prefactors.

Finally, note that if $G_{\bf 1}(x)$ is actually crossing invariant, that is, if all four marginal fields are identical, $G_{\bf 1}(x)=\langle \Phi(\infty)\Phi(1)\Phi(x)\Phi(0)\rangle$, then it satisfies the crossing relations
\be\label{1111}
G_{\bf 1}(x)=G_{\bf 1}(1-x)\ , \qquad G_{\bf 1}({\textstyle\frac1x}) =G_{\bf 1}(x)|x|^4\ .
\ee
This implies that $F_{\bf 1}(\tau)$ is actually invariant under the full modular group $\Gamma_1$.

Let us summarize here the action of all elements of the quotient group $S_3=\Gamma_1/\Gamma(2)$ on the various quantities that will play a role:
\begin{align}
	&\textrm{Label:}&&\quad\;\; A&&\quad\;\;\;B&&\quad\;\;\;C&&\quad\;\;\;D&&\quad\;\;\;E&&\quad\;\;\;F\\
	&\gamma:&&
	\begin{pmatrix} 1&0\\0&1\end{pmatrix}&&
	\begin{pmatrix} 0&-1\\1&0\end{pmatrix}&&
	\begin{pmatrix} 1&0\\-1&1\end{pmatrix}&&
	\begin{pmatrix} 1&-1\\0&1\end{pmatrix}&&
	\begin{pmatrix} 0&1\\-1&1\end{pmatrix}&&
	\begin{pmatrix} 1&-1\\1&0\end{pmatrix}\\
	&\lambda(\gamma\tau): &&\quad\;\;\lambda &&\quad 1-\lambda &&\quad\;\;\;\frac1\lambda
	&&\;\;\frac\lambda{(\lambda-1)}&&\;\;\frac1{(1-\lambda)}&&\;\;\;\frac{(\lambda-1)}\lambda\\
	&x(\gamma\tau): &&\quad\;\;x &&\quad 1-x&&\;\;\frac x{(x-1)} &&\quad\;\;\;\frac1x 
	&&\;\;\frac{(x-1)}x &&\;\;\frac1{(1-x)} \label{xtrafo} \\
	&\sigma_\gamma(1234): &&\;\;(1234) && \;\;\;(1432) && \;\;\;(2134) && \;\;\;(4231) && \;\;\;(2431) && \;\;\;(4132)\\
	&Z: &&\;\;\; Z_{\e_0,\e_1}&&\;\;\;\;Z_{\e_1,\e_0}&&\;\;Z_{\e_0+\e_1,\e_1}&&\;\;Z_{\e_0,\e_0+\e_1}
	&&\;\;\;Z_{\e_1,\e_0+\e_1}&&\;\;Z_{\e_0+\e_1,\e_0}\label{Ztrafo}
\end{align}
Here $ Z_{\e_0,\e_1}$ is the partition function of a single boson compactified on $S^1$, as defined in eq. (\ref{Z4pt}), which will show up in our computations in sections \ref{s:T16} and \ref{s:T8}. Note that we have $Z(\tau)= Z_{\gamma}(\gamma^{-1}\tau)$. We have included its transformation properties in the above table for completeness.

\section{Integrals on the cover}\label{s:Integral}

\subsection{Cover map and the integral}\label{ss:coverInt}

Let us now apply the insights of section~\ref{s:cover} to compute the integral of a 4-point function of four marginal operators. We will at first neglect all issues of regularization, and return to this in section~\ref{ss:regscheme}.

We want to express the integral (\ref{H0}) in coordinates of the cover.
Since $x(\tau)$ is a bijection between $\C$ and $\mathbb{H}_+/\Gamma(2)$, the integral over $x$ turns into an integral over the fundamental region of $\Gamma(2)$,
\be
\int_\C  G(x) d^2x=\int_{\mathbb{H}_+/\Gamma(2)}d^2\tau \left|\frac{\partial x}{\partial\tau}\right|^2G(\tau)
=\int_{\mathbb{H}_+/\Gamma(2)}y^{-2} d^2\tau F(\tau)
\ee
Here we use the function $F(\tau)$ which we introduced in (\ref{Fdef}), and $y^{-2} d^2\tau=\Im(\tau)^{-2} d\tau \wedge d\bar\tau$ is the usual $\Gamma_1$ invariant measure.
A fundamental domain $\cF_2:=\mathbb{H}_+/\Gamma(2)$ is plotted in figure~\ref{f:F2}. Note that $\cF_2$ has three cusps at $0,1,i\infty$,
which by eq. (\ref{cuspmap}) correspond to the poles of $G(x)$ at $x=0,\infty,1$, respectively. The cusps are thus exactly the points where the integrand $F(\tau)$ can diverge. 

To evaluate the integral, we want to expand $F(\tau)$ in $q$ and $\qb$,  integrate term by term, and then sum over all terms.
This procedure is however problematic because of the cusps on the real line: since $q^h \qb^\hb$ is not exponentially suppressed there, there is no reason to expect the sum over all terms to converge. 

To avoid these problematic cusps on the real line, we split the total integral over $\cF_2$ into six integrals over the six images fundamental domain $\cF_1$ displayed in figure~\ref{f:F2}, and then map those to the standard keyhole region plotted in blue using the known $\Gamma_1$ transformation properties of $F(\tau)$. This gives
\be
\int_\C  G(x) d^2x = \sum_{\gamma\in \Gamma_1/\Gamma(2)} 
\int_{\cF_1}y^{-2} d^2\tau F_{\sigma_\gamma}(\tau)
=\int_{\cF_1}y^{-2} d^2\tau \tilde F(\tau)
\ ,
\ee
where we have used (\ref{Ftrafo}) and defined
\be\label{Ftilde}
\tilde F(\tau):=  \sum_{\gamma\in \Gamma_1/\Gamma(2)} 
F_{\sigma_\gamma}(\tau)\ .
\ee
The advantage of this setup is that the two corners of $\cF_1$ are closest to the real axis and have values
\be
\tau_c = e^{\frac{\pi i}3}\ , \ e^{\frac{2\pi i}3}\ ,
\ee
so that in all of the integration region 
\be
|q| < e^{-\pi \sqrt{3}}\ .
\ee
We can thus expect that the sum over the integrated terms converges very quickly.
Note in particular that on the base, these corner points correspond to 
\be
x_c= \frac{1}{2}\pm i\frac{\sqrt{3}}{2}\ ,
\ee
with $|x_c|=1$.
This explains why it is useful to go to the cover to perform the $\tau$ integral order by order in $q$ rather than doing the same for $x$ integral: the radius of convergence of $G(x)$ around 0 is 1, and even if we use crossing symmetry, we need to integrate up to the points $x_c$, which have $|x_c|=1$. The expansion in $x$ will thus converge badly, whereas the expansion in $q$ converges much better.

Note that $\tilde F(\tau)$ defined in (\ref{Ftilde}) is now invariant under all of $\Gamma_1$, since it is the sum of all images of $\Gamma_1/\Gamma(2)$.
If the four marginal fields happen to be identical, then $F(\tau)$ is already invariant under $\Gamma_1$, so that $\tilde F(\tau)=6F(\tau)$ and
\be
\int_\C  G(x) d^2x = 6\int_{F_1}y^{-2} d^2\tau F(\tau)\ .
\ee
Next, we expand $\tilde F(\tau)$ around the cusp $i\infty$,
\be\label{Ftexp}
\tilde F(\tau)= \sum_{H,\Hb}y^2a_{H,\Hb} q^H \qb^{\Hb}
\ee
where $H-\bar H \in \Z$, since $\tilde F$ is invariant under $T$. 
To find the actual expression (\ref{Ftexp}), we compute the $q$-expansions of the $F_{\sigma}(\tau)$ by expanding the appropriate $G_\sigma(x)$ around 0.

We then integrate (\ref{Ftexp}) term by term.
To evaluate the integral over $\tau=x+iy$, we first switch to complex coordinates,
\be
\int_{\cF_1}y^{-2} dx dy \tilde F(\tau)=\frac{i}{2}\int_{\cF_1} d\tau d\bar\tau y^{-2}\tilde F(\tau)\ .
\ee
Let us also assume for the moment that $H+\bar H >0$ --- we will return to regularization issues momentarily. A given term can then be integrated as
\be
I(H,\Hb):=\frac{i}{2}\int_{\cF_1} e^{2\pi i(H\tau -\bar H \bar \tau)} d \tau d\bar \tau
=
\left\{ \begin{array}{cc}
	(4\pi\bar H)^{-1} \int_{\partial \cF_1} e^{2\pi i(H\tau -\bar H \bar \tau)}d \tau
	&: \bar H \neq 0\\
	\!\!\!\!\!\!\!\!\!\!\!\!(4\pi H)^{-1}\int_{\partial \cF_1} e^{2\pi iH\tau}d\bar \tau &: \bar H =0
\end{array}
\right.
\ee
where we have used Stokes' theorem. Due to invariance under $\tau \mapsto \tau+1$, the contributions of the two vertical sides of $\cF_1$ cancel out, and the integrand vanishes at $i\infty$.
The only contribution thus comes from the circle segment of the keyhole region running between $e^{\frac{\pi i}3}$ and $e^{\frac{2\pi i}3}$,
\be\label{IHHb}
I(H,\Hb)
=
\left\{ \begin{array}{cc}
	-\frac{i}{4\pi\bar H}\int_{\frac\pi3}^{\frac{2\pi}3} e^{2\pi i(He^{i\theta} -\Hb e^{-i\theta})} e^{i\theta}d\theta
	&: \bar H \neq 0\\
	\frac{i}{4\pi H}\int_{\frac\pi3}^{\frac{2\pi}3} e^{2\pi i(He^{i\theta} -\bar H e^{-i\theta})}e^{-i\theta}d\theta &: \bar H =0
\end{array}
\right.
\ee
The total integral is then given by 
\be\label{intsum}
\int_{\cF_1}y^{-2} dx dy \tilde F(\tau)
=\sum_{H,\bar H} a_{H,\bar H} I(H,\bar H)\ .
\ee
Note that $I(H,\bar H)$ is exponentially suppressed in $H$ and $\bar H$, so that the total integral, \ie the sum over all terms, converges fairly quickly. It may be possible to improve on the convergence by using methods described in \cite{2018arXiv180910908C}, but we did not attempt to do so. Instead, we simply evaluate the integrals $I(H,\bar H)$ in (\ref{IHHb}) numerically using Mathematica.

\subsection{Regularization scheme}\label{ss:regscheme}
In the above, we avoided any divergence issues at $x=0$ and $\infty$ by mapping the corresponding cusps to the cusp $\tau=i\infty$. 
To avoid divergences at that cusp, we simply assumed that the integrand decayed exponentially. There are of course terms which do not decay, which is why the integral needs to be regularized. This is what we will discuss now.

If there is a field of weight $h,\hb$ in the OPE of the two marginal fields at $x$ and $1$, then $G(x)$ has a term 
\be
G(x)\sim \frac{1}{(1-x)^{2-h}(1-\xb)^{2-\hb}} 
\ee
in its expansion around $x=1$. Using (\ref{x}) and (\ref{JacExp}), we find that on the cover, this translates to $F(\tau)$ expanding around $q=0$ as
\be
F(\tau)\sim y^2 q^{\frac h2-\frac12} \qb^{\frac{\hb}2-\frac12}\ .
\ee
From (\ref{Ftexp}), we see that $H={\textstyle\frac h2-\frac12}$ and $\Hb ={\textstyle\frac\hb2-\frac12}$. It is thus exactly the relevant fields which lead to exponential growth at $i\infty$, and which therefore must be regularized.

To regularize them, the usual $\epsilon$-disc scheme is not very convenient, since the discs will be mapped to more complicated shapes under the cover map. Instead, we will therefore use the subtraction scheme described in section~\ref{s:2ndorder}: namely, we subtract the contributions of relevant fields at the three singularities $0,1,\infty$. To this end we define the three regulators
\bea\label{G0reg}
G_0(x)&=& \sum_{\Delta_\phi<2} \frac{C_{\Phi_4\Phi_3\phi}C_{\Phi_2\Phi_1\phi}}{x^{2-h_\phi}\xb^{2-\hb_\phi}}\\
\label{G1reg}
G_1(x)&=& \sum_{\Delta_\phi<2} \frac{C_{\Phi_2\Phi_3\phi}C_{\Phi_1\Phi_4\phi}}{(1-x)^{2-h_\phi}(1-\xb)^{2-\hb_\phi}}\\
\label{Ginfreg}
G_\infty(x)&=& \sum_{\Delta_\phi<2} \frac{C_{\Phi_1\Phi_3\phi}C_{\Phi_2\Phi_4\phi}}{x^{h_\phi}\xb^{\hb_\phi}}
\eea
Note that here we assume that there are no field of total weight $\Delta=2$. (In fact, we only need to assume that there are no marginal fields in the OPE, since the integrals of (\ref{G0reg})--(\ref{Ginfreg}) do not give a constant term if the weight of $\varphi$ is not equal to (1,1). Since our marginal fields $\Phi_i$ are marginal at first order, this is automatically the case.) The regularized correlation function $\Greg(x)$ is the defined as
\be
\Greg(x)=G(x)-G_{reg}(x)= G(x)- G_0(x)-G_1(x)-G_\infty(x)\ 
\ee
and has no singularities when integrated over $\C$. Next we need to check  the $\epsilon$-disc regulated integral of the regulators
\be
\int_{\C-D_\epsilon} G_{0,1,\infty}(x)d^2x 
\ee
for constant terms, which would affect the shift as described below (\ref{Dint}). Clearly the only way to obtain a constant term is if $\Delta_\phi$ is integer. We therefore only need to check the cases $(0,0)$, $(\frac12,\frac12)$, $(1,0)$ and $(0,1)$. Explicit computation shows that only the last two cases give a constant term. However, for the examples we consider below, there are never such fields in the OPE. For the case where we have four identical marginal operators, this can be seen using the symmetry in exchanging the positions of two fields in the correlation function: the existence of vector fields in the OPE yield terms proportional to $\textstyle\frac1{z^2\bar z}$, which is, however, odd under $z\to-z$. {\footnote{We thank the referee for this comment.}} We can thus simply use (\ref{H0}) to compute the lifting.

We can now repeat the procedure described in section~\ref{ss:coverInt} using the regularized correlator. We define
\be\label{Fr}
F^{(r)}(\tau) := y^2 |x'|^2\Greg(x)\ ,
\ee
which is still $\Gamma(2)$ invariant, and then use the regulated $F^{(r)}(\tau)$, rather than $F(\tau)$, as the integrand. To obtain $\tilde F^{(r)}(\tau)$, we can write 
\be
F^{(r)}(\gamma\tau) = F_{\sigma_\gamma}(\tau) - F_{reg}(\gamma\tau)\ 
\ee
where 
\be
F_{reg}(\gamma\tau) = y^2 \left|\frac{\partial (\gamma x)}{\partial \tau}\right|^2 G_{reg}(\gamma x)\ ,
\ee
and we use the expressions (\ref{xtrafo}) for $\gamma x = x(\gamma\tau)$.
$\tilde F^{(r)}(\tau)$ is then simply the sum of these over $S_3$.
The total integral is thus simply
\be
\int_\C  \Greg(x) d^2x = \int_{\cF_1}y^{-2} d^2\tau  \tilde F^{(r)}(\tau)
\ee
which is no longer divergent at $i\infty$, and can be evaluated term by term using (\ref{intsum}).

\section{Marginal twisted fields: $\mathbb T^{16}/\Z_2$}\label{s:T16}

\subsection{Torus orbifolds}
Let us now investigate marginal fields of $\Z_2$ torus orbifolds. The $\Z_2$ symmetry with which we are orbifolding is the usual $X \mapsto -X$ symmetry of the torus. As usual, the orbifolding introduces twisted sectors, one for each fixed point of $\Z_2$. Perturbation theory in the untwisted sector is not very interesting, since the theory is free there. The twisted sectors, however, are more interesting.

Let us denote the ground state of the twisted sector corresponding to the fixed point $\varepsilon$ by $\sigma_\varepsilon$. If there are $c$ free bosons, the ground state has conformal weight
\be\label{groundstateweight}
h_{\sigma_\varepsilon}= \left(\frac c{16},\frac c{16}\right)\ .
\ee
States in the twisted sectors can then be constructed by acting with  descendants of the bosons $\partial X$, as long as we ensure to construct orbifold invariant states by including an even total number of descendants. Note that in the twisted sector the free bosons are half-integer moded,
\be
\partial X(z) = \sum_{r\in \Z+\frac12} \partial X_{r} z^{-r-1}\ ,
\ee
so that they indeed pick up a minus sign when rotated around the twist field $\sigma(0)$.

From (\ref{groundstateweight}) we see that there are two ways of obtaining a marginal field in the twisted sector: if $c=16$, then $\sigma$ by itself is already marginal. If $c=8$, then its bosonic descendant
\be\label{T8marginal}
\partial X_{-\frac12} \bar\partial X_{-\frac12} \sigma_\varepsilon
\ee
is a marginal field. In both cases the field continues to be marginal at first order, since the 3-point function $\langle \sigma \sigma \sigma \rangle$ vanishes for the twist field and all its descendants, as it contains an odd number of twist fields. The question that we want to study is therefore if the field continues to be marginal at second order. Since the theory is not supersymmetric, there is no reason to expect so, and indeed we will find that the field ceases to be marginal at second order, and is therefore not a moduli of the theory.

\subsection{$\mathbb T^1/\Z_2$}
Let us start out with a single boson compactified on a circle of radius $R$. There are then two fixed points at 0 and $\pi R$, which we label by $\e \in \Z_2$. The 4-point function of their ground states $\sigma_\e$ can be written as \cite{Dixon:1986qv}:
\be\label{4ptepsilon}
G_{\e_0,\e_1}(x):=
\langle \sigma_{0}(\infty)\sigma_{ \varepsilon_1}(1)
\sigma_{ \varepsilon_1+ \varepsilon_0}(x)
\sigma_{\varepsilon_0}(0) \rangle
=Z_{cov}Z_{\e_0,\e_1}(\tau)
\ee
where
\be\label{Z4pt}
Z_{\e_0,\e_1}(\tau)=\frac{1}{|\eta(\tau)|^2}\sum_{m\in \Z,n\in 2\Z+\e_0}(-1)^{m\e_1} q^{\frac14\left(\frac mR+\frac{nR}2\right)^2}
\qb^{\frac14\left(\frac mR-\frac{nR}2\right)^2}\ 
\ee
and
\be
Z_{cov}(\tau)=2^{-\frac23}|x(1-x)|^{-\frac1{12}}\ .
\ee

Let us discuss (\ref{4ptepsilon}) in detail. First of all, note that we can shift the fractional charge $\e$ by an overall constant, which allows us to set the charge of the first field to zero without loss of generality. Due to overall charge conservation modulo $\Z_2$, the charge of the third field is fixed. Eq. (\ref{4ptepsilon}) is thus the most general non-vanishing 4-point function. To compute (\ref{4ptepsilon}), we note that the four twist fields lead to two $\Z_2$ branch cuts. To eliminate those, we map the four-punctured sphere to its double cover, using the map introduced in section~\ref{s:cover}. The prefactor $Z_{cov}$ in  (\ref{4ptepsilon}) comes from this transformation \cite{Lunin:2000yv}. On the cover, the ground states become vacuum states, which explains why $Z_{\e_0,\e_1}(\tau)$ is essentially the 0-point function of the free boson on the covering torus.

Note however that there are a few crucial differences to the usual 0-point function. First of all, the sum is over even winding modes only. Second, there is an additional factor of 2 in the exponent due to the fact that the torus is a double cover of the original sphere. Last, only if $\e_0=\e_1=0$, that is only if all four twist fields are in the same fixed point sector, do we recover the usual partition function. In all other cases, the fractional charges of the various twist fields shift the winding modes, or introduce additional signs.
Finally note that $Z_{\e_0,\e_1}(\tau)$ transforms under $\Gamma_1$ as 
\be\label{Zmodtrafo}
Z_{\e_0,\e_1}(\tau+1)= Z_{\e_0,\e_0+\e_1}(\tau)\ , \qquad
Z_{\e_0,\e_1}(-\textstyle\frac1\tau)= Z_{\e_1,\e_0}(\tau)\ .
\ee
This is compatible with the the transformation properties listed in (\ref{Ztrafo}). In particular, $Z_{\e_0,\e_1}(\tau)$ is invariant under $T^2$ and $S T^2 S$, which means that it is invariant under $\Gamma(2)$, as is required.

Let us discuss the case of four identical fields in slightly more detail. If we have $\e_0=\e_1=0$,  $Z_{00}(\tau)$ is fully invariant under $\Gamma_1$. As expected, we get 
\be\label{crossing}
G_{00}(1-x)= G_{00}(x)\ ,
\ee
and also
\be
G_{00}(\textstyle\frac1x)= 2^{-\frac23}|x^{-3}x(x-1)|^{-\frac1{12}} Z_{00}(\tau)= |x|^{\frac14} G_{00}(x) = x^{2h}\bar x^{2\bar h} G_{00}(x)\ ,
\ee
with $h = \bar h = \frac{1}{16}$. To investigate the expansion around $x=0$, we can first use (\ref{crossing}) and then (\ref{qexpansion}) to find
\be
G_{00}(x)= |x|^{-\frac14}\sum_{h,\hb} \frac{1}{16^{h+\hb}}x^{h} \xb^{\hb}(1+\ldots)
\ee
where the sum is over 
\be
h =\textstyle\frac12 (\frac mR+\frac{nR}2)^2\ , \qquad \hb = \frac12(\frac mR-\frac{nR}2)^2\ , \qquad m\in\Z,n\in 2\Z\ .
\ee
The dots include all bosonic descendants of the lattice fields. This is compatible with the fact that the two ground states of weights $h=\bar h= \frac{1}{16}$ at $x$ and $0$ fuse to produce the vacuum and higher modes.

\subsection{$\mathbb T^{16}/\Z_2$}
Let us now consider the case of 16 free bosons on $\mathbb T^{16}$, so that $\sigma_\e$ is marginal. For simplicity, we take $\mathbb T$ to be the product of 16 $S^1$ of identical radius $R$. In total, there are now $2^{16}$ fixed points, which we denote by $\ve\in \Z_2^{16}$. The total correlation function can then be written as a product of 16 individual correlation functions. That is, we have
\be\label{GT16}
G_{\ve_0,\ve_1}(x)= 2^{-\frac{32}3}|x(1-x)|^{-\frac43} Z_{\ve_0,\ve_1}(\tau)
\ee
with
\be\label{ZT16}
Z_{\ve_0,\ve_1}(\tau) = \frac{1}{|\eta(\tau)|^{32}}\prod_{i=1}^{16} Z_{\e^{(i)}_0,\e^{(i)}_1}(\tau)
\ee
where $Z_{\e^{(i)}_0,\e^{(i)}_1}(\tau)$ is given by (\ref{Z4pt}). We then have
\be
F(\tau)= y^2 |x'|^2G_{\vec\e_0,\vec\e_1}(x)= y^2 |x'| 2^{-\frac{32}3}|x(1-x)|^{-\frac43}Z_{\vec\e_0,\vec\e_1}(\tau)\ .
\ee
As usual, the prefactor $y^2 |x'| 2^{-\frac{32}3}|x(1-x)|^{-\frac43}$ is invariant under modular transformations. To obtain $\tilde F(\tau)$, we use (\ref{Ztrafo}) for the sum over the crossing group. The unregularized integral is then
\begin{multline}\label{unregint}
\int_\C G_{\vec\e_0,\vec\e_1}(x)d^2x =\int_{\cF_1} d^2\tau |x'| 2^{-\frac{32}3}|x(1-x)|^{-\frac43}\times\\
\times\left(Z_{\ve_0,\ve_1}+Z_{\ve_1,\ve_0}+Z_{\ve_0+\ve_1,\ve_1}+Z_{\ve_1,\ve_0+\ve_1}+Z_{\ve_0+\ve_1,\ve_0}+Z_{\ve_0,\ve_0+\ve_1}\right)\ .
\end{multline}
We can write down the analogous expression for the regularized integral.

\subsection{The lifting matrix}
Let us now compute the lifting of the conformal weight of the marginal fields $\sigma_{\ve}$ when perturbing by $\Phi$. For concreteness, we will choose the radii of all tori to be $R=\frac23$. Without loss of generality we also choose our marginal field to be $\Phi=\sigma_0$. We want to compute the lifting matrix $M^0_{\vec\e_1\vec\e_2}$ for the $2^{16}$ marginal fields $\sigma_\e$. In particular, $M^0_{00}$ will gives us the lifting of $\Phi$, which will tell us whether $\Phi$ is marginal at second order or not. 

The lifting matrix to second order in $\Phi$ is thus given by
\be\label{Deps1eps2}
M^0_{\vec\e_1\vec\e_2} = \int_\C d^2x \Big(\langle\sigma_{\vec\e_1}(\infty)\Phi(1)\Phi(x)\sigma_{\vec\e_2}(0)\rangle - G_{reg}(x)\Big)\ .
\ee
Charge conservation immediately implies that $\vec\e_1=\vec\e_2=:\vec\e$, which means that $M^0$ is diagonal. To bring the 4-point function to the form (\ref{4ptepsilon}), we shift the overall charges  such that the correlator is now given by
\be\label{T16_4pf}
G_{0,\vec\e}(x)= \langle\sigma_{0}(\infty)\sigma_{\ve}(1)\sigma_{\ve}(x)\sigma_{0}(0)\rangle\ ,
\ee
that is, $\ve_1=\ve$, $\ve_0=0$.

For a start, let us compute the entry $M^0_{00}$, which will give us the lifting of the perturbing field $\Phi$ itself. The 4-point function then has four identical fields $\Phi$ and is therefore invariant under crossing transformation. We use (\ref{GT16}) to compute the regulator (\ref{G0reg}) and obtain
\begin{multline}
\qquad G_0 = \frac{1}{x^2 \xb^2}+\frac{20\ 2^{\frac49} }{3 x ^{\frac{14}9}\xb^{\frac59}}+\frac{2^{\frac{29}9} }{9 x^{\frac{16}9}\xb^{\frac79}}+
\frac{70\ 2^{\frac23}}{(x \xb)^{\frac43}}+\frac{20\ 2^{\frac49}}{3 x ^{\frac59}\xb^{\frac{14}9}}\\
+\frac{2^{\frac{29}9}}{9 x^{\frac79}\xb^{\frac{16}9}}+\frac{911}{{2}^{\frac{19}9} (x \xb)^{\frac{10}9}}+\frac{30\ 2^{\frac49}}{(x \xb)^{\frac{14}9}}+
\frac{2^{\frac{29}9}}{(x \xb)^{\frac{16}9}}\ .
\end{multline}
Because the correlator is crossing symmetric, we obtain the same coefficients for $G_1$ and $G_\infty$, allowing us to compute $G_{reg}(x)$. 
Putting everything together, we obtain $\tilde F_{reg}(\tau)$ from (\ref{Fr}), which we then expand in $q, \qb$. We then integrate it term by term up to a cutoff $H_{max}$ in $H,\Hb$. We have not tried to estimate the error, but the integral converges fairly well in $H_{max}$ as is shown in table \ref{tableT16}. Using these results together with (\ref{hshift2}), we find that $\sigma_0$ is not exactly marginal, but rather marginally relevant.

\begin{table}[h]
	\centering
\begin{tabular}{|c|cccc|}
	\hline
	$h,\hb\leq H_{max} =$ & 1 &2&3&4\\
	\hline
	$M^0_{00}$: & 15306.4  & 15364.9  & 15363.8  &15363.9 \\
	\hline
\end{tabular}
\caption{Anomalous dimension of the marginal field $\Phi=\sigma_{0}$ for $\mathbb T^{16}/\Z_2$ for four fields $\Phi$. This is associated with the element $M^0_{00}$ of the mixing matrix. $H_{max}$ is the upper cutoff of the integral and the results show that the integral converges well.}\label{tableT16}
\end{table}

Next we want to compute the lifting of arbitrary twist fields $\sigma_{\ve}$ under perturbations by $\Phi$. To that end we need to compute the lifting matrix $M^0_{\vec\e_1\vec\e_2}$ for general $\vec\e_{1,2}$. From charge conservation we immediately see that $\vec\e_1 =\vec\e_2$, so that $M^0$ is diagonal. 
From (\ref{ZT16}) we see that of the $2^{16}$ possible choices for $\vec\e$, only the number of non-zero entries of $\vec\e$ matters. We will denote the number of such entries by $n$ with $n=0,1,\ldots,16$. It is thus enough to compute only 17 different cases.

Now that the 4-point function is no longer crossing symmetric, we need to take into account the different images of the crossing group.
For a general 4-point function, from (\ref{unregint}) we see that we need six images of the crossing group in (\ref{Ztrafo}),
\begin{align}\label{ZAZBZC}
Z_A= Z_{\ve_0,\ve_1\ ,}&& Z_B=Z_{\ve_1,\ve_0}\ ,&&Z_C=Z_{\ve_0+\ve_1,\ve_1}\ ,\\
Z_D=Z_{\ve_0,\ve_0+\ve_1}\ ,&& Z_E=Z_{\ve_1,\ve_0+\ve_1}\ ,&&Z_F=Z_{\ve_0+\ve_1,\ve_0}\ .
\end{align}
In our case, we have in fact $\vec\e_0=0,\vec\e_1=\vec\e$, so that some of those functions are identical.
Similarly the regulators $G_{0,1,\infty}(x)$ will now all have different coefficients. To obtain these, we expand $G_{\ve_0,\ve_1}(x)$ around $x=0,1$, and $\infty$. In practice, we know the expansion of $Z$ around $q=0$, that is $\tau=i\infty$, from (\ref{ZT16}), which corresponds to the expansion around $x=1$. This means we can take
\be
G_1(x) = G_{\ve_0,\ve_1}(x)|_{\rm relevant}^{x=1}= 2^{-\frac{32}3}|x(1-x)|^{-\frac43} Z_A(\tau)|_{\rm relevant}^{\tau=i\infty}\ .
\ee
To find the expansion around $x=0$, we use the modular transformation properties of $x(\tau)$ and $Z_{\ve_0,\ve_1}(\tau)$. Namely, we have 
\begin{multline}
\qquad\qquad\qquad\qquad\; G_0(x)=G_{\ve_0,\ve_1}(x)|_{\rm relevant}^{x=0} = 2^{-\frac{32}3}|x(1-x)|^{-\frac43} Z_{\ve_0,\ve_1}(\tau)|_{\rm relevant}^{\tau=0} \\
=2^{-\frac{32}3}|x(1-x)|^{-\frac43}Z_B(-{\textstyle\frac1\tau})|^{-\frac1\tau=i\infty}_{\rm relevant}\ .
\end{multline}
which, because of (\ref{qtildeexpansion}), indeed gives an expansion in $x$ around 0.
Finally, from (\ref{Ztrafo}) we see that $Z_{\ve_0,\ve_1}(\tau) = Z_{\ve_0+\ve_1,\ve_0}(-{\textstyle\frac1{(\tau-1)}})$, which we use to obtain
\begin{multline}
\qquad\qquad\quad G_\infty(x)=G_{\ve_0,\ve_1}(x)|_{\rm relevant}^{x=\infty}=2^{-\frac{32}3}|x(1-x)|^{-\frac43} Z_{\ve_0,\ve_1}(\tau)|_{\rm relevant}^{\tau=1} \\
=2^{-\frac{32}3}|x(1-x)|^{-\frac43}Z_F(-{\textstyle\frac1{(\tau-1)}})|^{-{\textstyle\frac1{(\tau-1)}}=i\infty}_{\rm relevant}\ ,
\end{multline}
which, because of $q(-{\textstyle\frac1{(\tau-1)}})= {\textstyle\frac1{256}}\,x^{-2}+\ldots$, indeed gives an expansion in $x$ around infinity. This way we obtain
\be
G_{reg}(x)=G_0(x)+G_1(x)+G_\infty(x)\ 
\ee
and finally 
\be
F^{(r)}(\tau)= F(\tau)- y^2 |x'|^2 G_{reg}(\tau)\ .
\ee
The total integral (\ref{Deps1eps2}) is then given by
\begin{multline}
\qquad\quad\int_\C  \Greg(x) d^2x = 
\int_{\cF_1}y^{-2} d^2\tau \sum_{\sigma\in S_3}  F^{(r)}_{\sigma}(\tau)\\
=\int_{\cF_1} y^{-2}d^2\tau \left(F^{(r)}_{\e_0,\e_1}+F^{(r)}_{\e_1,\e_0}+F^{(r)}_{\e_0+\e_1,\e_1}+F^{(r)}_{\e_1,\e_0+\e_1}
+F^{(r)}_{\e_0+\e_1,\e_0}+F^{(r)}_{\e_0,\e_0+\e_1}\right)\ .
\end{multline}
In practice, to obtain $F^{(r)}_{\sigma}(\tau)$, for the $F_\sigma(\tau)$ part we simply use (\ref{Ftrafo}) and the fact that $\sigma$ acts on $\ve_{0,1}$ as described in (\ref{Ztrafo}) . For the regulators, as described in section~\ref{ss:regscheme}, we find it easier to explicitly insert the transformed variable $\gamma x =x(\gamma\tau)$ into $|x'|^2 G_{tot}(x)$, and then expand in $q$.

From this we compute the lifting matrix by integrating the expansion of $\tilde F^{(r)}$ up to $H_{max}=4$.
We list the entries of the lifting matrix in table \ref{T16matrix}, where $n$ denotes the number of non-zero entries of $\ve$. We find that the integrals at that order have converged to the precision given in the table. Overall, the lifting matrix is diagonal, and all diagonal entries are non-vanishing. It follows that the marginal operator $\Phi$, and therefore, all marginal operators $\sigma_\e$, are not exactly marginal. Their shift in conformal weight under perturbation by $\Phi$ is given by
\be
h_n(\lambda) = 1 - \frac{\pi}2M^0_{nn} \lambda^2 +O(\lambda^3)\ .
\ee

\begin{table}[h]
	\centering
	\begin{tabular}{|ccccccccc|}
		\hline
$n:$&0&1&2&3&4&5&6&7\\
\hline
$M^0_{nn}:$&15364&-552 &5724&-525 &-161&142&183&49\\
\hline
8&9&10&11&12&13&14&15&16\\
\hline
-106&-150&-30&192&301&-99&-1606&-5049&-11501 \\
\hline
	\end{tabular}
	\caption{Entries of the mixing matrix $M^0_{\vec\e_1\vec\e_2}$ (\ref{Deps1eps2}). Here $M^0_{nn}$ gives the shift in weight of the marginal field $\Phi=\sigma_{\ve}$ under the perturbation $\Phi= \sigma_0$, and $n$ denotes the numbers of non-zero entries in $\ve$. Because of charge conservation there are only 17 distinct values out of the $2^{16}$ indices --- see the discussion above (\ref{ZAZBZC}).}\label{T16matrix}
\end{table}

\section{Marginal twisted fields: $\mathbb T^{8}/\Z_2$}\label{s:T8}

\subsection{Torus correlators of descendant fields}
Here we want to investigate the lifting of marginal fields of the form $\partial X_{-\frac12}\bar \partial X_{-\frac12}\sigma$. To compute their correlation functions, we proceed again by mapping to the double cover. This means that we need to compute correlation functions of bosonic descendants on the torus.

Our starting point for this is the torus correlation function for bosonic operators of the form
\be
O_{\alpha_i}(z,\bar z)= e^{i\sqrt{2}\alpha_i X(z,\bar z)}\ .
\ee
This operator has conformal weight $h_i=\bar h_i=\alpha_i^2$. The correlation function of $n$ such operators on the torus may, for instance, be found in \cite{DHoker:1988pdl} for the uncompactified case and in \cite{DiFrancesco:1997nk} for the compactified case,\footnote{Note that there is a typo in eq. (12.150) of \cite{DiFrancesco:1997nk} which we have fixed in eq. (\ref{torusnpoint}) above.} and is given by
\begin{multline}\label{torusnpoint}
Z(z_i)=\langle O_{\alpha_1}(z_1,\zb_1)\ldots O_{\alpha_n}(z_n,\zb_n)\rangle \\
= \prod_{i<j} \left(\frac{\partial_z \theta_1(0|\tau)}{\theta_1(z_{ij}|\tau)} \right)^{-2\alpha_i\alpha_j} 
\overline{ \left(\frac{\partial_z \theta_1(0|\tau)}{\theta_1(z_{ij}|\tau)} \right)}^{-2\alphab_i\alphab_j}
\frac1{|\eta(\tau)|^2}\sum_{\alpha,\alphab} q^h \bar q^{\bar h} e^{4i\pi(\sum_k \alpha \alpha_k z_k - \bar \alpha \bar \alpha_k \bar z_k )}\ .
\end{multline}
Here we  choose the $\alpha_k$ such that charge conservation is satisfied, and the sum is over the internal left- and right-moving $\alpha, \alphab$.

To obtain torus correlation function of fields $\partial X$ from this, we use
\be
\partial_z X = \lim_{\alpha\to0} \frac{\partial_z O_\alpha}{i\sqrt2\alpha}
\ee
and take limits of derivatives of eq. (\ref{torusnpoint}). To compute 2-point functions for instance, we start with
\be
\langle O_{\alpha_1}(z_1,\zb_1) O_{-\alpha_1}(z_2,\zb_2)\rangle 
= \left|\frac{\theta_1({z_1-z_2}|\tau)}{\partial_z \theta_1(0|\tau)} \right|^{4h_1} 
\frac1{|\eta(\tau)|^2}\sum_{\alpha,\alphab} q^h \bar q^{\bar h} e^{4i\pi(\alpha \alpha z_{12} - \bar \alpha \bar \alpha \bar z_{12} )}
\ee
Taking the derivative $\partial_1\partial_2$ pulls down various factors of $h$ and $\alpha$. Because of $\alpha\to0$, we only need to keep terms quadratic in $\alpha$ or linear in $h$. This means that we will either have to act with both derivatives on the $\theta$ functions, or with both derivatives on the sum. In total, we obtain
\be\label{hol2pt}
\langle \partial X(z_1)\partial X(z_2) \rangle= \sum_{\alpha,\alphab} Z_{\alpha,\alphab}
\ee
where
\be
Z_{\alpha,\alphab}= \left(K(z_1-z_2)
- 8\pi^2 \alpha^2 \right)\frac1{|\eta(\tau)|^2}q^{\alpha^2} \bar q^{\alphab^2}\ .
\ee
Here we have defined the Green's function
\be
K(z)= \frac{\theta''_1(z|\tau)\theta_1(z|\tau)-(\theta'_1(z|\tau))^2}{\theta_1(z|\tau)^2}\ .
\ee

It is straightforward to generalize this to more complicated correlation functions. In general, $Z_{\alpha,\alphab}$ factorizes into a left-moving and a right-moving part. For the 2-point function with both left- and right-moving descendants
\be\label{2pfdxdxb_i}
\langle (\partial X\bar\partial X)(z_1,\zb_1)\;(\partial X\bar\partial X)(z_2,\zb_2) \rangle
\ee 
we find 
\be\label{2pfdxdxb_ii}
Z_{\alpha,\alphab}=Z_\alpha \bar Z_\alphab
\ee
with 
\be\label{2pfdxdxb_iii}
Z_\alpha = \left(K(z_1-z_2)
- 8\pi^2 \alpha^2 \right)\frac1{\eta(\tau)}q^{\alpha^2}\ .
\ee
The general structure of a $n$-point correlation function on a torus is the following:
\be\label{Ztorus}
Z(z_n) = \sum_{\alpha,\alphab} Z_{\alpha,\alphab}(z_n)\ .
\ee
Here $Z_\alpha$ is the `quantum part' of the correlation function around a given classical solution of momentum $\alpha,\alphab$, and the total correlation function is the sum over all classical solutions.

Let us now turn to the correlation function that we need for the computation of the lifting matrix. It has 4 fields $\partial X \bar \partial X$ located at the four branch points, that is, at $z_1=\frac\tau2, z_2 = \frac12+\frac\tau2, z_3 = 0$, and $z_4=\frac12$. We can repeat the procedure above by taking a total of eight derivatives. The resulting expression, expanded in $q$, gives
\be
Z_\alpha = \frac1{\eta(\tau)} q^{\alpha^2}
\Big(\left(64 \pi ^4 \alpha ^4-16 \pi ^4 \alpha ^2+\pi ^4\right)+\left(384 \pi ^4 \alpha ^2+144 \pi ^4\right) q + \ldots \Big)
\ee
and similarly for $\bar Z_\alphab$. Multiplying holomorphic and anti-holomorphic parts together, we obtain
\begin{multline}\label{Z4ptdesc}
\;Z_{\alpha,\alphab} = \frac1{|\eta(\tau)|^2} q^{\alpha^2} \qb^{\alphab^2}
\Big(\left(64 \pi ^4 \alpha ^4-16 \pi ^4 \alpha ^2+\pi ^4\right)+\left(384 \pi ^4 \alpha ^2+144 \pi ^4\right) q + \ldots \Big)\times\\
\times\Big(\left(64 \pi ^4 \alphab ^4-16 \pi ^4 \alphab ^2+\pi ^4\right)+\left(384 \pi ^4 \alphab ^2+144 \pi ^4\right) \qb + \ldots \Big)\ .
\end{multline}
We now need to convert these torus correlation functions to twist field correlation functions on the base space.

\subsection{Twisted fields and the cover map}
So far we only discussed correlation functions of twist ground states $\sigma$.
Let us now discuss how to compute correlation functions of general fields in the twisted sector. Suppose $\phi$ is a $\Z_2$ invariant field of weight $(h,\hb)$. Let $\tilde\sigma$ be a twisted field in the OPE of $\phi$ with $\sigma$,
\be\label{tildesigmaOPE}
\phi(z,\zb)\sigma(0) \sim \frac1{z^{h_\sigma+h-h_{\tilde\sigma}}\zb^{\hb_\sigma+\hb-\hb_{\tilde\sigma}}}\tilde\sigma(0) +\ldots\ ,
\ee
where $\sigma$ has dimensions $h_\sigma=\hb_{\sigma}$. We claim that the leading field in the OPE has dimensions $h_{\tilde\sigma}=\hb_{\tilde\sigma}=h_\sigma+\frac h2$. To see this, we start out with the correlation function on the cover
\be
Z(t_i)\ ,
\ee
where the $t_i$ are the cover coordinates of the various fields $\phi_i$. We will send the $t_i$ to the images of the branching points on the torus to recover the fields $\tilde \sigma$. Note, in particular, that on the torus this limit is not divergent, other than possible normal ordering prescriptions between different $\phi_i$ at the same point.

On the base the correlation function is then
\be\label{Gbase}
G(z_i) = \prod_i (f'(t_i))^{-h_i}(\bar f'(t_i))^{-\hb_i} Z(t_i) Z_\sigma
\ee
where $f(t)$ is the cover map such that $z_i=f(t_i)$, and $Z_\sigma$ is the contribution of the cover map coming from the twist fields \cite{Lunin:2000yv}. Assume that the branch point of $f(t)$ at $t=0$ is given by
\be\label{zeqbt2}
z(t)=bt^2 +\ldots\ .
\ee
As pointed out above, when sending $t\to0$, $Z(t)$ remains regular. We do however pick up a singularity of the form
\be
\frac{2^{-h-\hb}b^{-\frac h2}\bar b^{-\frac{\hb}2}}{z^{\frac h2}\zb^{\frac{\hb}2}}
\ee
in eq. (\ref{Gbase}) from the prefactors $f'$. On the base, the interpretation of this singularity is simply the appearance of $\tilde \sigma$ with the claimed dimensions in the OPE (\ref{tildesigmaOPE}). In total, the correlation function of $\tilde \sigma$ is thus
\be\label{tildesigmaOPE_ii}
\langle \tilde \sigma(0)\ldots \rangle = \lim_{t\to 0} z^{\frac h2}\zb^{\frac{\hb}2} Z(z_i)= 2^{-h-\hb}b^{-\frac h2}\bar b^{-\frac{\hb}2} Z(t_i) Z_\sigma\ .
\ee
This generalizes the expressions derived in \cite{Lunin:2000yv} to non-holomorphic fields. Note that we fixed the normalization of the twisted field $\tilde\sigma$ through (\ref{tildesigmaOPE}). It may therefore differ from more standard normalizations.

In the case at hand, we want to take $\phi = (\partial X \bar \partial X)(z,\bar z)$ in eq. (\ref{tildesigmaOPE}). We want to work with $\phi$ directly, since it is invariant under $\Z_2$, unlike its holomorphic and anti-holomorphic components $\partial X$ and $\bar \partial X$. We then define the marginal field $\Phi$ by
\be\label{phiOPEnorm}
\phi(z,\bar z)\sigma(0)=\frac14 \,\frac1{z^{\frac12}\zb^{\frac12}}\,\Phi(0)+\ldots\ ,
\ee
where we introduced the factor of $\frac14$ to ensure that $\Phi$ will be normalized correctly. To put it another way, up to normalization, $\Phi$ is given by $\partial X_{-\frac12}\bar \partial X_{-\frac12}\sigma$.

To check the normalization of $\Phi$, let us compute the 2-point function $\langle\Phi(a)\Phi(0)\rangle$. Following \cite{Lunin:2000yv}, we use the cover map
\be\label{zt2covermap}
f(t)= \frac{at^2}{2t-1}\ ,
\ee
which maps the insertion points to 0 and 1 on the spherical cover. The 2-point function on the cover is thus simply $|t_1-t_2|^{-4}$ which then gives 1 when we send $t_1\to 0$ and $t_2\to 1$. The cover map itself gives $b=a$ for both branch points, see eq. (\ref{zeqbt2}). Using the fact that $Z_\sigma$ is simply the 2-point function $\langle\sigma(a)\sigma(0)\rangle= |a|^{-4h_\sigma}$, we find
\be
\langle\Phi(a)\Phi(0)\rangle = \frac1{|a|^{4(h_\sigma+\frac12)}}\ ,
\ee
which shows that with our convention $\Phi$ is indeed normalized to 1.

\subsection{Correlators of twisted fields}
Let us start with the correlation function for a single free boson of two twist fields and two moduli inserted at $1$ and $x$, 
\be
\langle \sigma(\infty)\Phi(1)\Phi(x)\sigma(0) \rangle\ .
\ee
This means we have $t(z=1)=\frac\tau2$ and $t(z=x)=\frac12+\frac\tau2$, giving $b_1=e_2-e_3= -\pi^2\theta_2^4$ and $b_x=\frac{(e_3-e_1)(e_3-e_2)}{(e_2-e_1)} = \pi^2 \theta_2^4\theta_4^4\theta_3^{-4}$, respectively. The contribution for four twist fields $\sigma$ is again given by
\be\label{baretwist}
Z_\sigma(x)= 2^{-\frac23}|x(1-x)|^{-\frac1{12}}\ .
\ee
Putting together eqs. (\ref{2pfdxdxb_i})-(\ref{2pfdxdxb_i}), (\ref{Gbase}), and (\ref{baretwist}), we find
\begin{multline}
\langle\sigma(\infty)\Phi(1)\Phi(x)\sigma(0)\rangle=2^{-\frac23}|x(1-x)|^{-\frac1{12}}|\pi^{-2}\theta_4^{-2}\theta_2^{-4}\theta_3^2|^2\times\\
\times \sum_{\alpha,\alphab}\left(K(\textstyle\frac12)- 8\pi^2 \alpha^2 \right)\left(\overline{K(\textstyle\frac12)}- 8\pi^2 \alphab^2 \right)
\frac1{|\eta(\tau)|^2}q^{\alpha^2}q^{\alphab^2}\ .
\end{multline}
Following \cite{Dixon:1986qv}, we are summing over
\be
\alpha=\frac12\Big(mR+\frac n{2R}\Big)\qquad \alphab=\frac12\Big(mR-\frac n{2R}\Big)
\ee
with $m \in \Z$, $n \in 2\Z$, so that $n$ is even. Expanding around $x=1$ we find
\be
\langle\sigma(\infty)\Phi(1)\Phi(x)\sigma(0)\rangle\big|_{x=1}=\frac 1{(1-x)^{\frac98}(1-\xb)^{\frac98}} + \ldots
\ee
which agrees with the fact that $\Phi$ has weight $(\textstyle\frac9{16},\textstyle\frac9{16})$.  

We can now finally put everything together to compute the shift in weight for the marginal field $\Phi$.
We will not compute the full lifting matrix, but instead concentrate on a single modulus 
$\Phi$ whose 4-point function we compute. For concreteness, we choose
\be\label{T8phi}
\Phi \sim \partial X^1_{-\frac12} \bar \partial X^1_{-\frac12} \sigma_0\ ,
\ee
where the normalization is understood to be fixed by (\ref{phiOPEnorm}). The 4-point function $\langle \Phi(\infty)\Phi(1)\Phi(x)\Phi(0)\rangle$ is then computed by taking (\ref{Ztorus}) together with (\ref{Z4ptdesc}) and multiplying with seven factors of the torus partition function. We then multiply it with the appropriate cover map contribution, see eq. (\ref{Gbase}):
\be
G(x) = Z_{cov}Z\ .
\ee
The cover map contribution $Z_{cov}$ consists of the contribution from the twist fields and from the other fields. To go down to the sphere, we now have $b_0=e_3-e_1 = -\pi^2\theta_4^4$, $b_x=\frac{(e_3-e_1)(e_3-e_2)}{e_2-e_1}=\pi^2 \theta_2^4\theta_4^4\theta_3^{-4}$,
$b_1=e_2-e_3= -\pi^2\theta_2^4$, and $b_\infty= (e_2-e_1)^{-1}=-\pi^{-2}\theta^{-4}_3$. 
Putting everything together we get
\be
Z_{cov} = \left|\pi^{-4} \theta^{-4}_2\theta^{-4}_4 2^{-\frac83}(x(1-x))^{-\frac13}\right|^2
\ee

Evaluating the integral with Mathematica, we evaluate the mixing matrix element $M^0_{00}$ (\ref{Deps1eps2}). Table \ref{tableT8} shows the anomalous dimensions for 5 values of the upper cutoff $H_{max}$ and we again find that the integral converges very well. This establishes that $\Phi$ is not marginal at second order, but becomes irrelevant. 

\begin{table}[h!]
	\centering
	\begin{tabular}{|c|ccccc|}
		\hline
		$h,\hb\leq H_{max} =$ & 1 &2&3&4&5\\
		\hline
		$M^0$ & -113.209 & -101.395 & -101.85 &-101.84&-101.841\\
		\hline
	\end{tabular}
\caption{anomalous dimension of the marginal field $\Phi$ (\ref{T8phi}) for the theory of free bosons on $\mathbb T^{8}$. All the four marginal fields in the 4-point function (\ref{Deps1eps2}) sit at the same fixed point, which corresponds to the element $M^0_{00}$ of the mixing matrix. $H_{max}$ is the upper cutoff of the integral as before, and the integral converges well.}\label{tableT8}
\end{table}

\section*{Acknowledgments}
We are happy to thank Eric D'Hoker, Matthias Gaberdiel, Daniel Friedan, Hirosi Ooguri, and Martin Raum for helpful discussions and Nathan Benjamin for detailed comments on the draft. The authors were supported by the Swiss National Science Foundation through the NCCR SwissMAP for part of this research.


\bibliographystyle{utphys}
\bibliography{refmain}

\end{document}